\documentclass[aps,prb,reprint,groupedaddress]{revtex4-1}
\usepackage[colorlinks]{hyperref}
\usepackage{graphicx}
\usepackage{dcolumn}
\usepackage{amsmath}
\usepackage{lipsum}
\usepackage{amssymb}
\usepackage{amsthm}
\usepackage{mathtools}
\usepackage{bm}

\begin{document}
\title{Gate induced monolayer behavior in  twisted bilayer black phosphorus}

\author{Cem Sevik}
\affiliation{Department of Mechanical Engineering, Faculty of Engineering, Anadolu University, Eskisehir, TR 26555, Turkey}
\email{csevik@anadolu.edu.tr}

\author{John R. Wallbank}
\affiliation{National Graphene Institute, University of Manchester, Manchester, M13 9PL, UK}
\email{john.wallbank@manchester.ac.uk}

\author{O\u{g}uz G\"{u}lseren}
\affiliation{Department of Physics, Bilkent University, Bilkent, Ankara 06800, Turkey}
\email{gulseren@fen.bilkent.edu.tr}

\author{Fran\c{c}ois M. Peeters}
\affiliation{National Graphene Institute, University of Manchester, Manchester, M13 9PL, UK}
\affiliation{Department of Physics, University of Antwerp, Groenenborgerlaan 171, B-2020, Antwerpen, Belgium}
\email{francois.peeters@uantwerpen.be}

\author{Deniz \c{C}ak{\i}r}
\affiliation{Department of Physics, University of Antwerp, Groenenborgerlaan 171, B-2020, Antwerpen, Belgium} 
\affiliation{Department of Physics and Astrophysics, University of North Dakota, Grand Forks, North Dakota 58202, USA}
\email{deniz.cakir@und.edu}

\date{\today}

\begin{abstract}
Optical and electronic properties of black phosphorus strongly depend on the number of layers and type of stacking. Using first-principles calculations within the framework of density functional theory, we investigate the electronic properties of bilayer black phosphorus with an interlayer twist angle of 90$^\circ$.  These calculations are complemented with a simple $\vec{k}\cdot\vec{p}$ model which is able to capture most of the low energy features and is valid for arbitrary twist angles.   
The electronic spectrum of 90$^\circ$ twisted bilayer black phosphorus is found to be x-y isotropic in contrast to the monolayer.  However x-y anisotropy, and a partial return to monolayer-like behavior, particularly in the valence band, can be induced by an external out-of-plane electric field. 
Moreover, the preferred hole effective mass can be rotated by 90$^\circ$ simply by changing the direction of the applied electric field. In particular, a $+$0.4 ($-$0.4) V/{\AA} out-of-plane electric field results in a $\sim$60\% increase in the hole effective mass along the $\mathbf{y}$ ($\mathbf{x}$) axis and enhances the $m^*_{\mathbf{y}}/m^*_{\mathbf{x}}$ ($m^*_{\mathbf{x}}/m^*_{\mathbf{y}}$) ratio as much as by a factor of 40. Our DFT and $\vec{k}\cdot\vec{p}$ simulations clearly indicate that the twist angle in combination with an appropriate gate voltage is a novel way to tune the electronic and optical properties of bilayer phosphorus and it gives us a new degree of freedom to engineer the properties of black phosphorus based devices.
\end{abstract}
\maketitle

\section{Introduction}\label{intro}
Within the family of atomically thin two dimensional (2D) materials, black phosphorus occupies a special status because of its buckled structure~\cite{makale1-10}, its highly anisotropic transport properties and its intermediate size direct size bandgap~\cite{cakir1,cakir2,cakir3,makale1-12,makale1-13,makale2-11, makale2-12, makale2-13, makale2-14, makale2-15, makale3-3, makale3-6}. Experimental and theoretical studies showed that few-layer black phosphorus displays high and anisotropic carrier mobility~\cite{makale3-7, makale3-8}, excellent electron-channel contacts~\cite{makale1-10}, and electronic and optical properties that are tuned by the stacking order\cite{cakir3,makale4-9, makale4-10}, in-plane strain\cite{cakir1}, and external electric field~\cite{cakir1,makale2-11, makale4-11, makale4-12, makale4-13}. Meanwhile, high performance field-effect transistors based on black phosphorus have been successfully fabricated with a carrier mobility up to 1000~cm$^{2}$/V/s~\cite{makale1-10} and an ON/OFF ratio up to 10$^4$ at room temperature~\cite{makale1-11}. Moreover, black phosphorus has also been implemented into various electronic device applications including gas sensors~\cite{makale1-14}, $p-n$ junctions~\cite{makale1-15}, and solar cells~\cite{makale1-16}. The reported results show that black phosphorus has potential to be adopted in future technological applications. 

First principles calculations have recently indicated that the distinct potential of this material might be substantially enhanced. For instance, by demonstrating that the optical and electronic properties of bilayer black phosphorous are strongly influenced by the type of stacking. In particular, tunable band gap~\cite{makale1-16,makaleST-3,makaleST-2,cakir1}, carrier effective masses along different crystallographic directions\cite{makaleST-2, makaleST-3,cakir3}, and high solar power conversion efficiency have been distinctly demonstrated~\cite{makale1-16}. Moreover, considerable change in carrier mobilities~\cite{makaleEF-1,makale1-16} and a continuous transition from a normal insulator to a topological insulator and eventually to a metal as a function of external electric field applied along the out-of-plane direction have been substantiated as well~\cite{makaleEF-1}. 

In the present work, the electronic properties of bilayer black phosphorus with an interlayer twist angle of 90$^\circ$ are systematically investigated by using first-principles calculations within the framework of density functional theory. These results are complemented with an analytic $\vec{k}\cdot\vec{p}$ model that is applicable for arbitrary twist angle. 
First, the optimum stacking formation map is determined by considering a 20$\times$20 grid (in step of 0.25 {\AA}) of different possible arrangements. Then, the electronic properties including effective masses along crystallographic directions (few meV cohesive energy difference) are systematically investigated under the effect of applied out-of-plane external electric field for both minimum and non-minimum energy stackings. 

\section{Computational Method}
The simulations are performed within the framework of density functional theory (DFT) as implemented in the Vienna $Ab$-initio Simulation Package (VASP)~\cite{VASP1,VASP2,VASP3,VASP4}. The generalized gradient approximation (GGA) formalism~\cite{GGA1} is employed for the exchange-correlation potential. The projector augmented wave (PAW) method~\cite{PAW1,PAW2} and a plane-wave basis set with an energy cutoff of 400~eV are used in the calculations. A regular 1$\times$1$\times$1 $k$-mesh within the Monkhorst-Pack scheme~\cite{Pack1976} is adopted for Brillouin-zone integration. 
We employ  a 5$\times$7 supercell resulting in a nearly commensurate lattice with a lattice mismatch of $<$ 1\%  between top and bottom layers.
A vacuum spacing of at least 20~{\AA} is introduced between isolated bilayers to prevent spurious interaction. By using the conjugate gradient method, atomic positions and lattice constants are optimized until the Hellmann-Feynman forces are less than 0.01~eV/{\AA} and pressure on the supercells is decreased to values less than 1~kB. The van der Waals interaction between individual layers are taken into account\cite{vdw,dftd3} for a correct description of the structural and electronic properties. In order to investigate the equilibrium bilayer structure, formation energies of all the possible non-symmetric structures are calculated. The structural and electronic properties  are obtained in the presence of a perpendicular uniform electric field, ranging from -0.4 to 0.4~V/{\AA} in steps of 0.1~V/{\AA}. The influence of the applied electric field on the effective masses of carriers are investigated using the following formula for the mass tensor,
\begin{equation}
m^{*}_{i,j}=\hbar^{2}\left(\frac{\partial^{2}E}{\partial k_i\partial k_j}\right)^{-1}.
\end{equation}
where $\hbar$ is the reduced Planck constant, $k_i$ ($k_j$) is the wavevector along the $i$ ($j$) direction, and $E$ is the energy eigenvalue. Here, the $i$ and $j$ directions are used for the $\Gamma$-X and $\Gamma$-Y directions in the first Brillouin zone of bilayer black phosphorus. In order to obtain a reliable numerical estimate of the second derivative of the band energies, we use a dense two-dimensional $k$-point grid centered at the $\Gamma$ point. Numerical second derivatives are obtained by using four point forward and backward differences.

\section{Results}
\subsection{DFT results for 90$^\circ$ twisted angle }
\begin{figure}[h!]
 \centering
 \includegraphics[scale=0.4]{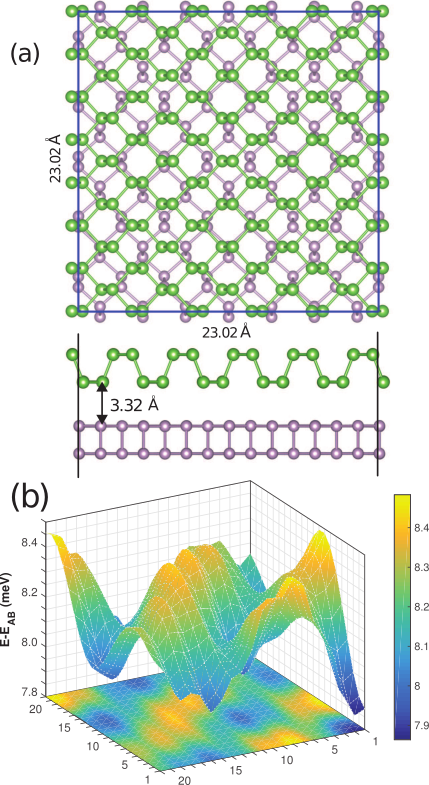}
 \caption{(a) Top and side views of 90$^\circ$ twisted bilayer black phosphorus, and (b) potential energy surface with respect to naturally stacked bilayer as a 
 function of lateral displacements of layers with respect to each other. }
 \label{energy}
\end{figure}

First of all, we investigate the structural and electronic properties of 90$^\circ$ twisted bilayer black phosphorus at zero electric field.  
Figure~\ref{energy} shows the variation of the formation energy difference between 90$^\circ$ twisted and AB stacked bilayer black phosphorous as a function of stacking of the rotated bilayer structure. Here, all the possible stacking arrangements are considered and the atomic positions are only relaxed along the $z$-direction.
In the rest of this work, unless otherwise stated, we focus on the lowest energy structure and relax the atomic positions in the in-plane directions as well.
From Fig. \ref{energy} we note that the energy difference between different stacking types of 90$^\circ$ twisted bilayer black phosphorus is at most 0.2~meV/atom. This means that layers can easily slide over each other. The van der Waals corrected interlayer separation of the 90$^\circ$ twisted bilayer black phosphorus is 3.32~{\AA} for the lowest energy structure, which is about 0.15~{\AA} larger than that of AB stacked bilayer\cite{cakir3}. The P-P bond lengths show only subtle changes in twisted bilayer as compared to the naturally stacked bilayer.

\begin{figure*}
 \centering
  \includegraphics[scale=0.3]{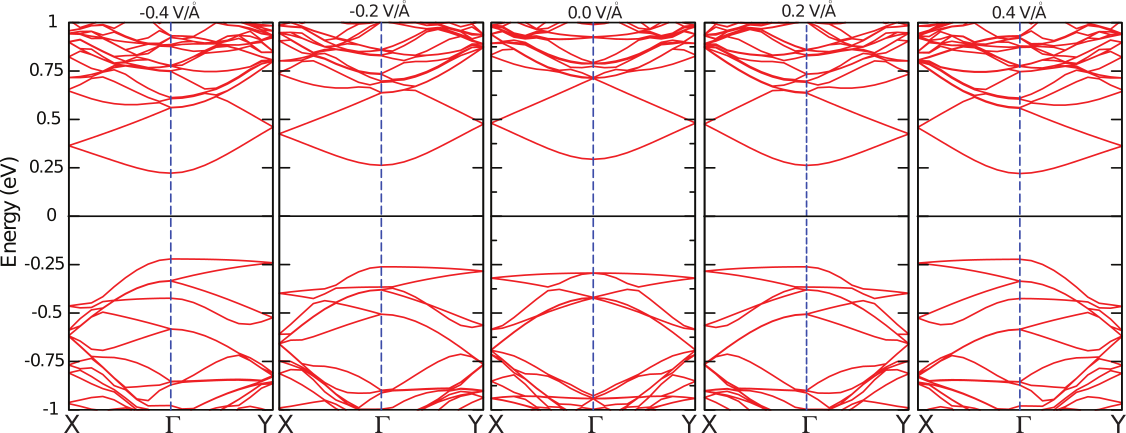}
  \caption{Band structure of 90$^\circ$ twisted bilayer black phosphorus for different values of the out-of-plane electric field.}
  \label{band}
\end{figure*}

Figure~\ref{band} displays the band structure of twisted bilayer black phosphorus as a function of applied electric field strength. 
In contrast to the AB stacked bilayer, 90$^\circ$ twisted bilayer has an x-y isotropic electronic structure at zero applied field, meaning that the energy bands in the $\Gamma$-X and $\Gamma$-Y directions are symmetric.
While the conduction band minimum (CBM) is  singly degenerate, the valence band maximum (VBM) is found to be almost doubly degenerate with an energy splitting of 0.1~meV.  

\begin{figure}[h!]
 \centering
  \includegraphics[scale=0.3]{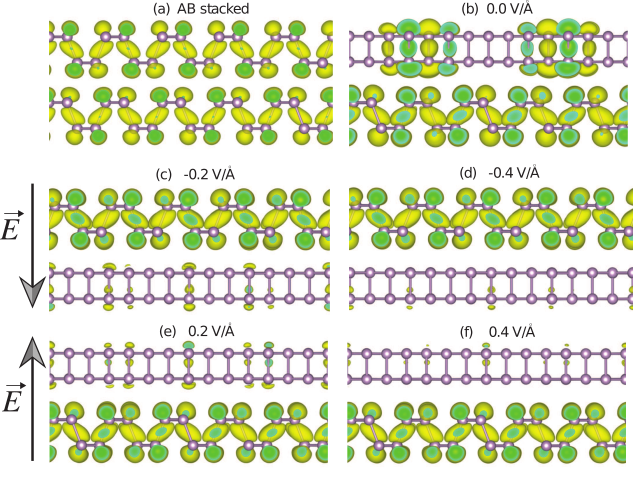}
  \caption{(b)-(f) The decomposed charge densities corresponding to the VBM of the 90$^\circ$ twisted bilayer phosphorus at the $\Gamma$ point for different values of the applied electric field. In (a), we also present the charge of VBM of the AB stacked bilayer black phosphorus.}
  \label{chg}
\end{figure}

For a better insight, Fig.~\ref{chg} depicts the absolute square of the wavefunctions corresponding to the top of the valence band. Naturally stacked (i.e. AB stacked) and 90$^\circ$ twisted bilayer black phosphorus display very different spatial characters at the VBM. In the case of AB stacked bilayer (Fig.~\ref{chg} a), the VBM wavefunction is equally localized over the two layers. In the 90$^\circ$ twisted bilayer (Fig.~\ref{chg} b-f), however, the wavefunction is mainly localized either on the bottom or top layer.  According to our calculations, the top of the valence band is mostly located on the bottom layer for zero field. The VBM is almost doubly degenerate with the other state 0.1 meV lower in energy. In that state, the wavefunction is mostly localized over the top layer. This spatial character of wavefunction especially implies the lack of a significant hybridization between states of individual layers near the VBM. Whereas, (not shown), the states of individual layers around CBM are strongly coupled, resulting in a wavefunction that is extended over both layers.

Next, we turn to investigate the response of the electronic structure of 90$^\circ$ twisted bilayer black phosphorus to a static out-of-plane electric field. The electronic properties of black phosphorus are very sensitive to in-plane/out-plane strain\cite{cakir1,makale4-13} and out-of-plane applied electric field\cite{makaleEF-1,makale4-11}. For all applied electric field values, the nature of the band gap remains direct at the $\Gamma$-point, see Fig. \ref{band}. In contrast to the zero field case,  the applied electric field restores the x-y anisotropic band structure of naturally stacked bilayer. In addition, the electric field lifts the degeneracy at the top of the valence band. Positive (negative) electric field (which is perpendicular to bilayer) shifts the bands along the $\Gamma$-Y  ($\Gamma$-X)-directions towards lower energies. The amount of splitting of doubly degenerate bands with electric field is found to be 0.1 eV for 0.2 V/{\AA} and becomes 0.2 eV for 0.4 V/{\AA}. 
After the degeneracy is lifted at the VBM, the state becomes localized on the layer with the higher potential. The applied field fully decouples the states of the two individual layers at the VBM, which can be easily seen by inspecting the charge density plots in Fig.~\ref{chg}. For high fields, the VBM is completely localized over either top or bottom layer depending on the sign or direction of the electric field.

By shifting constituent layers on top of each other, we can tune the interaction strength between the layers in AB stacked bilayer black phosphorus. 
This can be understood by calculating the charge density. 
Distinct from 90$^\circ$ twisted bilayer, different stacking types in the non-rotated bilayer 
result in different $\pi$-$\pi$ interaction distances and strengths, allowing the band gap to be tuned by  0.2-0.3 eV \cite{cakir3}. However, we find the band gap of 90$^\circ$ twisted bilayer is almost insensitive to the precise stacking, retaining a value of $\sim$0.58 eV for different stacking types. For some high energy stacking types, we predict that the energy splitting between the almost degenerate VBM states becomes 3 meV due to the reduction in symmetry and the change in interaction strength.

\begin{figure}[h!]
 \centering
 \includegraphics[scale=0.44]{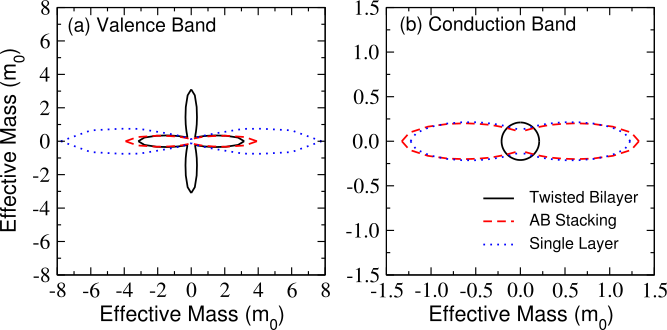}
 \caption{Direction dependent (a) hole and (b) electron masses in monolayer, AB stacked and 90$^\circ$ twisted bilayer black phosphorus. Each data point represents the end
point of a vector whose amplitude corresponds to the effective mass
in units of $m_0$ and the direction of this vector corresponds to the
direction in k space along which the mass is calculated.  }
 \label{mass-ab}
\end{figure}

\begin{figure}[h!]
 \centering
 \includegraphics[scale=0.44]{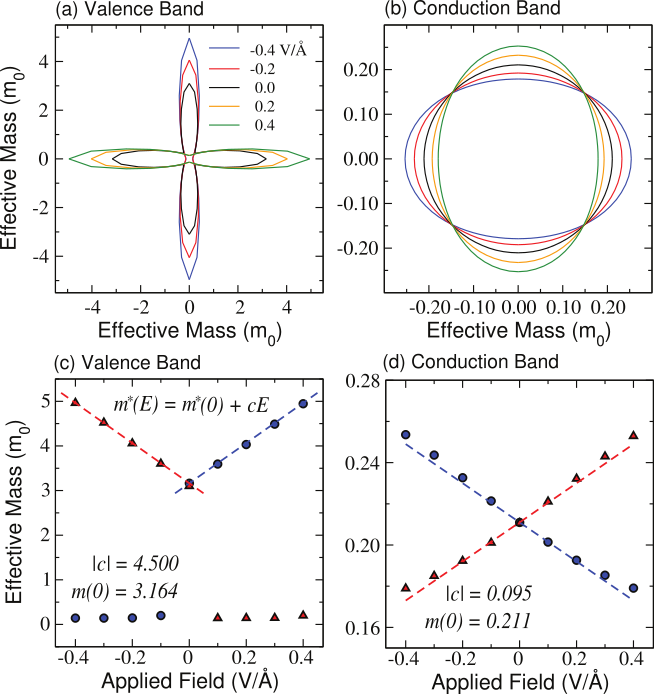}
 \caption{Direction-dependent (a) hole and (b) electron effective masses.  Each data point represents the end point of a vector whose amplitude corresponds to the effective mass in units of $m_0$ and the direction of this vector corresponds to the direction in k space along which the mass is calculated. Variation of (c) hole and (d) electron effective masses along the $\Gamma$-X (blue solid dot symbols)  and $\Gamma$-Y (red solid triangle symbols) directions as a function of out of plane electric field. Parameter $m^*$ is in units of $m_0$,
 $|c|$ is in units of $m_0$\AA/V.}
 \label{mass}
\end{figure}

The band edge effective masses along all reciprocal space directions (rather than just the $\Gamma$-X and $\Gamma$-Y directions) are displayed in Fig.~\ref{mass-ab} for AB stacked and 90$^\circ$ twisted bilayer black phosphorus. 
Notice that, these two different stacking types exhibit quite different electronic and transport properties.  First of all, due to its anisotropic band structure, the AB stacked black phosphorus has an apparent anisotropic hole and electron effective masses.  Rotation of  top layer with respect to bottom layer by 90$^\circ$ removes the anisotropy between the $x$ and $y$ directions. While the hole effective mass is 0.2 $m_0$ along the $\Gamma$-Y direction (armchair direction) in AB-stacked bilayer, it becomes 3.16  $m_0$ along the same direction in the 90$^\circ$ twisted bilayer. For comparison, we also show the effective masses for monolayer phosphorus which exhibits the strongest anisotropy.

We next investigate how an out-of-plane electric field changes the effective masses in 90$^\circ$ twisted bilayer, see Fig.~\ref{mass}. 
At zero field, the hole and electron effective masses are respectively 3.16 $m_0$ and 0.095 $m_0$, along both the $\Gamma$-X  and $\Gamma$-Y directions.
For holes, the application of a small positive (negative) electric field results in a sudden reduction of the hole effective mass along the 
the $\Gamma$-Y ($\Gamma$-X) direction, due to the lifting of the double degeneracy at the VBM and the partial return to monolayer behavior evinced by the wavefunction's localization on a single layer (Fig.~\ref{chg} c-f). 
For example, a +(-)0.1 V/{\AA} electric field results in a 0.2 $m_0$ effective mass along $\Gamma$-Y ($\Gamma$-X).
Hence the preferred direction of hole transport can switched by 90$^\circ$ when changing the direction of the electric field.  
In addition, the larger hole effective mass can be increased by up to 60\% by varying the applied field from 0 V/{\AA} to 0.4 V/{\AA}. 
This behavior of larger hole effective mass is well described by the linear dependence $m^*(E)$=$m^*$(0)+$cE$ where $m^*(E)$ and $m^*(0)$ are the hole effective masses in the presence and absence of the electric field, and fitting constant $c$ is given in Fig.~\ref{mass} (a).
The electron effective mass (Fig.~\ref{mass} (b,d)) also becomes anisotropic between $\Gamma$-X and $\Gamma$-Y directions when a finite electric field is applied, but the effect is weaker than for the holes, and does not display the discontinuity at zero electric field, due to the strong interlayer hybridization for this band. Indeed, the electron effective mass fall within 0.21$\pm$0.04 $m_0$ for the entire range of tested electric field strengths. Similar to the hole case, the variation of effective mass with electric field is well described by a linear fit (Fig.~\ref{mass} (b)).  
We also note that the interlayer separation of 90$^\circ$ twisted bilayer black phosphorus increases from 3.3 {\AA} at zero electric field to 3.4 {\AA} for a field strength of $\pm$ 0.4 V/{\AA}.

In this work, we employed semi-local functionals in order to investigate the electronic properties of twisted bilayer black phosphorus. We did not
consider GW corrections.  These corrections will likely change the results in a quantitative level, but we do not  expect any change in a
qualitative level. GW corrections will significantly enlarge band gap values. This is because of the fact we previously showed that many body
effects are significant for black phosphorus. For instance, the band gap for monolayer black phosphorus is
about 0.9 eV for PBE functional. It becomes 2.3 eV when we include GW corrections\cite{cakir1}. Regardless of
computational method (PBE, hybrid or GW) used in the calculations, the evolution of band gaps as a
function of number of layers has almost the same trend\cite{cakir1}. Similarly, band dispersions are not affected much. Therefore, we believe that
our calculated trends are correct.

\subsection{$\vec{k}\cdot\vec{p}$ approach for arbitrary twist angle}
We will now describe a simple $\vec{k}\cdot\vec{p}$ Hamiltonian which both reproduces the above described bandstructure features, particularly those near the band edges, and is able to describe arbitrarily misaligned twisted black phosphorous bilayers and their optical absorption properties. The  $\vec{k}\cdot\vec{p}$ Hamiltonian is written as,
\begin{align}\label{eq:H_kp}
 H=\begin{pmatrix}    H_{TT} & H_{TB}\\H_{TB}^\dagger & H_{BB} \end{pmatrix},
\end{align}
in a basis $(\psi^\text{T}_\text{C},\psi^\text{T}_\text{V},\psi^\text{B}_\text{C},\psi^\text{B}_\text{V})$ of $\Gamma$-point conduction/valence (C/V) band wavefunctions on the top/bottom (T/B) layers. For the intralayer terms \cite{rudenko-2014,ezawa-2014} we use,
\begin{widetext}
\begin{align}
 H_{\text{TT/BB}}=\begin{pmatrix}
                           \epsilon^0_\text{C}+u_\text{T/B} + \alpha^x_\text{C}\hat p_{ x}^2 + \alpha^y_\text{C}\hat p_{ y}^2  
                           &\gamma_{\text{ML}}\hat p_{ x} \\
                          \gamma_{\text{ML}}\hat p_{ x}
                          &\epsilon^0_\text{V}+u_\text{T/B} + \alpha^x_\text{V}\hat p_{ x}^2 + \alpha^y_\text{V}\hat p_{ y}^2 
\end{pmatrix},\nonumber
\end{align}
\end{widetext}
where $\hat{\vec{p}}=(-i\,\hbar\partial_x,-i\,\hbar\partial_y)$ for the top layer, $\hat{\vec{p}}=(i\,\hbar\partial_y,-i\,\hbar\partial_x)$ for the 90$^\circ$ rotated bottom layer, and $u_\text{T}=-u_\text{B}=  eEd/2$ for an out-of-plane electric field. 
For the interlayer hops we use
\begin{align}
 H_\text{T/B}=\begin{pmatrix}
                          \gamma_\text{C}  & 0                          \\
                          0                & \gamma_\text{V} \sum_{\delta\!\vec{g} =(\pm\delta\!g,0),(0,\pm\delta\!g )}   e^{i\delta\!\vec{g} \cdot \vec{r}}\\
                         \end{pmatrix},
\end{align}
where $\delta\!g=\frac{2\pi}{b}-\frac{2\pi}{a}$, $a/b$ are the $x/y$-directions lattice constants of monolayer black phosphorus, and $\gamma_\text{C/V}$ are hopping integrals. 
To obtain this Hamiltonian we approximated the interlayer hopping matrix elements using the overlap of the wavefunctions, $\langle \psi^\text{T}_\text{C/V}|H|\psi^\text{B}_\text{C/V} \rangle\sim  \langle \psi^\text{T}_\text{C/V}|\psi^\text{B}_\text{C/V} \rangle$, and expanded the $\Gamma$-point wavefunctions in the shortest few plane waves consistent with the translational and point-group symmetry of monolayer black phosphorus. We also neglected an interlayer valence-to-conduction band coupling due to its negligible affect on the band structure.

\begin{figure}[h!]
 \centering
 \includegraphics[scale=0.27]{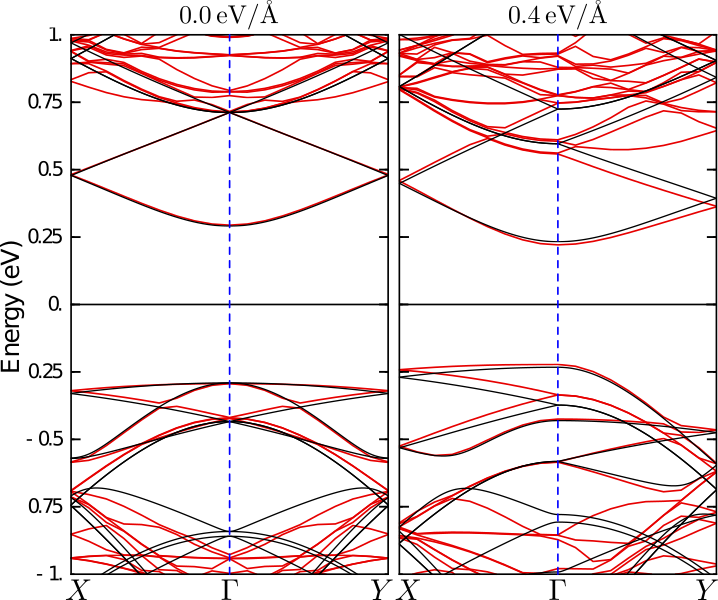}
 \caption{The bandstructure of 90$^\circ$ twisted black phosphorus calculated using either DFT (red), or the $\vec{k} \cdot \vec{p}$ model (black) for zero and finite electric fields. The values for the fitted parameters are: 
   $\gamma_\text{ML}=4.5\,$eV, $\gamma_\text{C}=0.39\,$eV, $\gamma_\text{V}=0.07\,$eV, $\epsilon^0_\text{C}-\epsilon^0_\text{V}=0.99\,$eV,  $\alpha^x_\text{C}=1.2\,$eV\AA$^{2}$, $\alpha^y_\text{C}=2.7\,$eV\AA$^{2}$, $\alpha^x_\text{V}=-5.9\,$eV\AA$^{2}$,   $\alpha^y_\text{V}=-2.0\,$eV\AA$^{2}$, $d=0.54\,\text{\AA}$.
 }
 \label{kdotp_vs_dft}
\end{figure}

The periodicity, $2\pi/\delta\!g $, which enters Hamiltonian \eqref{eq:H_kp} is that of the moir\'e pattern formed from the two black phosphorous layers, rather than the twice larger periodicity of the commensurate unit cell displayed in Fig.~\ref{energy}. Consequently, this Hamiltonian describes a twisted bilayer with a general incommensurate ratio between lattice constants $a$ and $b$, rather than the exact commensurate ratio $a/b=5/7$ used in the DFT calculations. 
Nevertheless, we fold the band structure into the smaller Brillouin zone of the commensurate unit cell to match the DFT bands, and fit all free parameters to obtain a good match (Fig.~\ref{kdotp_vs_dft} caption).

For zero applied electric field, Hamiltonian \eqref{eq:H_kp} posses a $C_{4v}$ symmetry  in which both the $\pi/4$-rotation and reflection in the $x\pm y=0$ planes are combined with the exchange of the top and bottom layers. 
Using this symmetry we find that the pair of doubly degenerate VBM states at the $\Gamma$-point belong to the two-dimensional E irreducible representation, while the singly degenerate CBM state belongs to the one dimensional B$_1$ irreducible representation (a "bonding'' state with the corresponding "anti-bonding'' state higher in energy by $2\gamma_\text{C}$).
The application of an electric field (Fig.~\ref{kdotp_vs_dft}, right panel) breaks the degeneracy of the VBM states so that the top of the valence band resembles that of the single isolated monolayer.
Note that the above discussed $C_{4v}$ symmetry is slightly broken by the commensuration of the lattices displayed in Fig.~\ref{energy}, which accounts for the slight lifting of the VBM degeneracy in the DFT calculated bandstructure.

\begin{figure}[h!]
 \centering
 \includegraphics[scale=0.2]{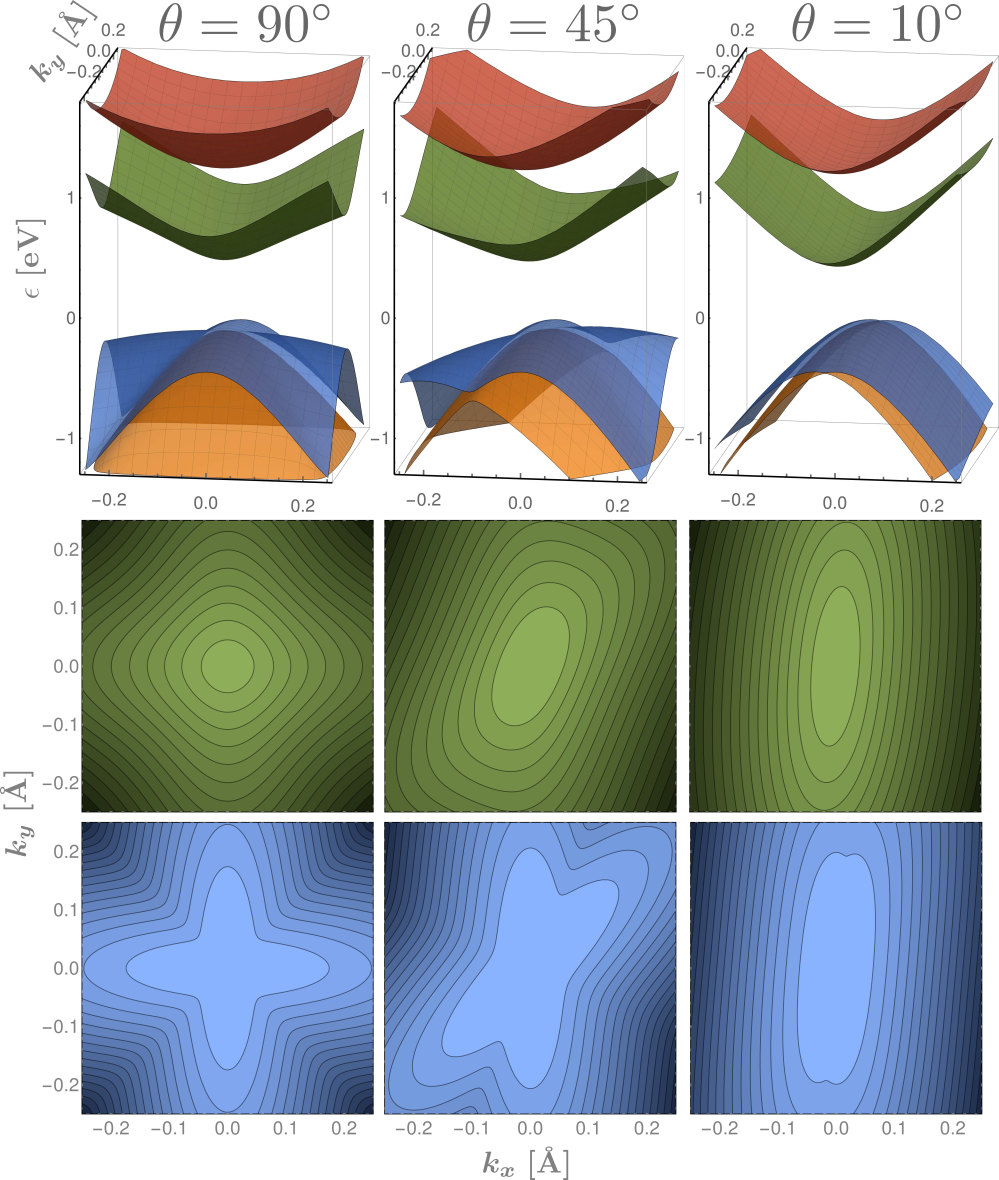}
 \caption{The low-energy dispersion surfaces of bilayer black phosphorus for different interlayer twist angles, $\theta$, (top row), and the corresponding contour plots of the CBM and VBM bands (middle and bottom rows respectively). Parameters as Fig.~\ref{kdotp_vs_dft} except here $\gamma_\text{V}=0$. }
 \label{kdotp_bands_contours}
\end{figure}

We can also adapt Hamiltonian \eqref{eq:H_kp} to the study of bilayer black phosphorous with an $\textit{arbitrary twist angle}$, $\theta$, by simply using the appropriately rotated momentum $\hat{\vec{p}}=(-i\,\hbar\partial_x\cos(\theta)+i\,\hbar\partial_y\sin(\theta) , -i\,\hbar\partial_x\sin(\theta)-i\,\hbar\partial_y\cos(\theta) )$ in $H_{BB}$ and neglecting the interlayer $\gamma_V$ hop. This approach is valid for 
wavevectors  $k< \text{min}\{ \delta\!g/2 , \theta \pi/a \}$ 
where both the $\gamma_V$ coupling and the superlattice effects, expected for an almost aligned ($\theta\ll1)$ bilayer, are irrelevant. Dispersion surfaces calculated using this method are displayed in Fig.~\ref{kdotp_bands_contours} for different choices of $\theta$. In this simple model varying $\theta$ breaks the four-fold rotational symmetry seen in the bandstructure but does not change the energies of the $\Gamma$-point band energies. Nevertheless the $\Gamma$-point degeneracy of the VBM is not protected by symmetry $\theta\neq90^\circ$, and would be slightly lifted if the neglected $\gamma_V$ hop or superlattice effects were included.

\begin{figure}[h!]
 \centering
 \includegraphics[scale=0.17]{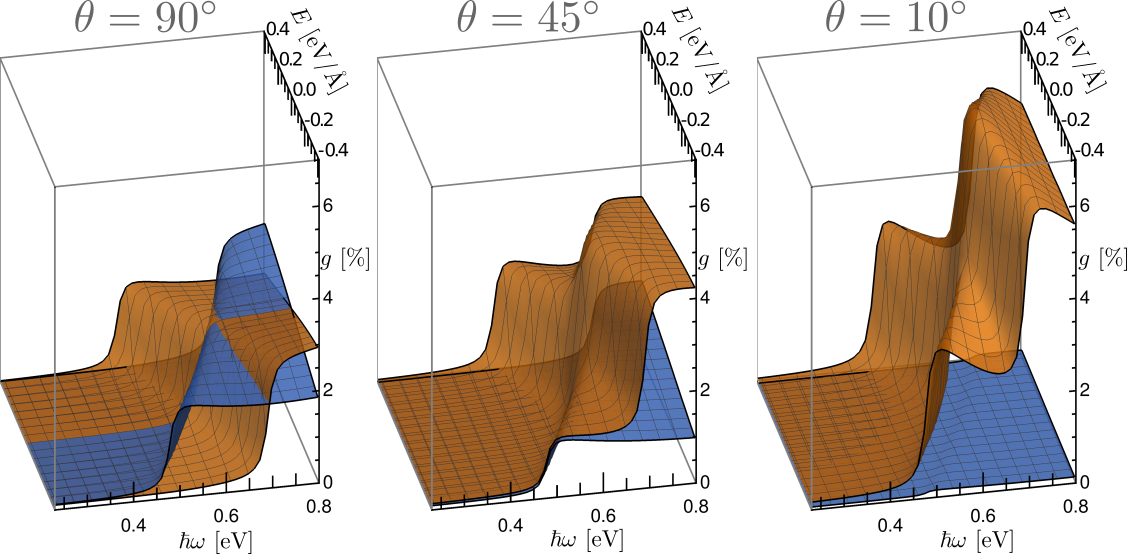}
 \caption{
 The optical absorption, $g$, of bilayer black phosphorous as a function of the applied electric field, $E$, and excitation energy, $\omega$, calculated for light polarized in the x/y-directions (orange/blue surfaces) and various interlayer twist angle. Band parameters are as per Fig.~\ref{kdotp_bands_contours}.
 }
 \label{optical_absorption}
\end{figure}

As an example of the effect of controllably breaking the $C_{4v}$ symmetry of Hamiltonian \eqref{eq:H_kp}, either by applying an electric field, or using an interlayer rotation $\theta\neq90^\circ$, we now calculate the optical absorption of linearly polarized light (Fig.~\ref{optical_absorption}).
Here we use the matrix element for optical transition $\sim\langle\psi_i|v_{j}|\psi_f\rangle$ with velocity operators $v_{j=x/y}=\partial H/\partial p{_j}$, introduce a phenomenological energy broadening $\eta=10\,$meV, and neglect many-body effects such as the formation of excitons.
For small interlayer twist angles (e.g.~$\theta=10^\circ$, right panel) the optical absorption resembles that of two isolated monolayers, displaying a strong preference to absorb light polarized in the $x$-direction, and step-like increases of the absorption when the excitation frequency allows electronic transitions between the VBM on either layer (which are split by the electric field) and the CBM. 
In contrast, a $90^\circ$ twist between the layers (left panel) and zero applied electric field produces an isotropic optical absorption.
However splitting the degeneracy of the VBM bands on the two layers produces a controllable linear dichroism in which the electric field direction selects the preferred polarization direction for absorption. We note that for intermediate twist angles (e.g. $\theta=45^\circ$, center panel) the absorption resembles a combination of the two extremes.

\section{Conclusions}
Using first principles and $\vec{k}\cdot\vec{p}$ based calculations, we investigated the structural, electronic and transport properties of 90$^\circ$ twisted bilayer black phosphorus. Even though twisted bilayer phosphorus displays isotropic electronic and transport properties, an out-of-plane electric field is able to create significant anisotropy in these properties. We demonstrated that the hole effective mass increases from 3.16 $m_0$ to 5 $m_0$ when the field is raised  from 0 V/{\AA} to 0.4 V/{\AA}, corresponding to a 60\% increase.  The hole effective mass approaches the value of the monolayer limit as the applied field increases and bilayer black phosphorus starts to display monolayer behavior for holes. The states near the VBM are localized over either top or bottom layer depending on the direction of the electric field. In addition, we predicted that the degeneracy of the highest occupied valence bands on the two layers splits by the applied electric field and this produces a controllable linear dichroism in which the electric field direction selects the preferred polarization direction for absorption. In summary, our calculations show that twisting combined with an appropriate  gate voltage gives us a new degree of freedom in manipulating the electronic, transport, optical and even thermoelectric properties of few-layer black phosphorus. For instance,  when the applied field is zero, the isotropic electronic and thermal properties give rise to isotropic thermoelectric properties in 90$^\circ$ twisted bilayer black phosphorus.  However, by the help of an out-of-plane applied electric field, we can tune the ratio of the Seebeck coefficients along the $\mathbf{x}$ and $\mathbf{y}$ directions. Similarly, the optical response can be changed from isotropic to anisotropic, which can be used in conceptually new device designs.

\section{acknowledgement}
This work was supported by the bilateral project between the The Scientific and Technological Research Council of Turkey (TUBITAK) and FWO-Flanders, Flemish Science Foundation (FWO-Vl) and the Methusalem foundation of the Flemish government. Computational resources were provided by TUBITAK ULAKBIM, High Performance and Grid Computing Center (TRGrid e-Infrastructure), and HPC infrastructure of the University of Antwerp (CalcUA) a division of the Flemish Supercomputer Center (VSC), which is funded by the Hercules foundation. We acknowledge the support from TUBITAK (Grant No. 115F024), ERC Synergy grant Hetero2D and the
EU Graphene Flagship Project. We also thank  Vladimir Fal'ko for helpful discussions.  

\bibliography{2d} 

\end{document}